# A GRADUATE COURSE ON E-COMMERCE INFORMATION SYSTEMS ENGINEERING

John Grundy

Department of Computer Science, University of Auckland
Private Bag 92019, Auckland, New Zealand
john-g@cs.auckland.ac.nz

Phone: +64-9-3737-599 ext 8761
Fax: +64-9-3737-453

**ABSTRACT**

Interest in developing and deploying E-commerce systems has grown greatly over the past few years, leading to a major shortage of qualified developers. This paper describes the development of a graduate course in E-commerce education that has been implemented as part of an Honours degree programme and a first year Masters offering for a traditional Information Systems programme. The aim of this course is to give graduate students a broad grounding and holistic set of skills to enable them to begin successful E-commerce systems development. This paper provides a background and context for this course, an overview and discussion of the course content, delivery methods and practical focus, and discusses experience in running and evolving the course over the past two years. We hope that our experiences and approaches will be useful for others who are planning similar courses focusing on E-commerce systems engineering.

## 1. INTRODUCTION

Many organisations are developing Information Systems to support E-commerce applications, typically using the World Wide Web to provide customer access to product and service purchase, and internet capabilities to support inter-organisational information exchange (Bambury, 1998; Bolin, 1998). There is thus a crucial need for Information Systems developers with skills in developing such applications. Such developers need a good understanding of both intra- and inter-organisational use of E-commerce systems, appropriate analysis and design techniques, and implementation, quality assurance and deployment technologies and methods.

This need for Information systems developers with a range of E-commerce system development skills has led us to design and run a graduate Information Systems course focusing on E-commerce systems development, as part of an Information Systems degree programme. This course can be taken either as an Honours year course (i.e. last year of 4-year professional degree) or as a first year Masters degree course. The course is aimed at equipping students, who have studied conventional IS analysis, design and implementation courses, with a range of skills and experience suitable for E-commerce systems development. The course takes a holistic approach, covering current IS development processes; E-commerce system IT policy, characteristics and deployment issues; requirements engineering; E-commerce system architectures and design techniques; user interface, database and middleware implementation technologies; and distributed IS quality assurance. The course uses a range of teaching strategies, including lectures, readings and web resources, student presentations, tutorial-based group activities and simulations, focused practical exercises, group development projects and individual research projects. This mix is intended to give students a balance of theory, practical development techniques, and practical system development experience for E-commerce systems engineering.

The following section provides a discussion of E-commerce system development issues we believe are important for students. We then discuss the educational context in which our graduate-level course is run, give an overview of the course structure, content and teaching philosophy and approaches. The following sections respectively discuss the mix of theory, practical development technique and practical project-based experience the course provides students, and our assessment strategies. We conclude with a discussion of the effectiveness of the course and lessons learned from its delivery and evolution.

## 2. BACKGROUND

There has been a tremendous growth in recent years in the use of "E-commerce", or "e-business", activity, and a corresponding growth in the need for Information Systems engineers with appropriate skills to realise such applications (Bolin, 1998; Dhamija *et al*., 1999; Jutla *et al*. 1999b; Ge & Sun, 2000). Much of the interest in E-commerce systems has focused on their perceived ability to provide better customer service via the WWW and more timely exchange of information between organisations via the internet, along with cost savings, improved organisational effectiveness and quicker response to market needs (Bambury, 1998; Bolin, 1998; Jutla *et al*. 1999b; Timmers, 1999). Many organisations have developed E-commerce systems of one kind or another, some being electronic forms of traditional business activities (such as order processing and customer liaison and purchasing) (Jutla *et al*. 1999a), while others are novel electronic business ventures (such as electronic product sales and virtual organisations) (Bambury, 1998; Fielding *et al*., 1998).

When developing E-commerce systems, a combination of management, technical and quality assurance issues must be addressed. Management and organisational issues include the use of appropriate project planning and cost/benefit analysis techniques, the choice of most appropriate development processes and tools, and user requirements being elicited and documented accurately (Schlueter & Shaw, 1997; Shafer *et al*., 1999). Many novel challenges are presented by E-commerce systems development, in addition to traditional Information Systems development issues. These include cost/benefit analysis for systems for untried application domains, highly distributed process and project management strategies (e.g. for "virtual" organisations), and the complexity of analysing requirements for a very diverse user base, which may not be easy to identify or may change after system deployment (Schlueter & Shaw, 1997).

Many challenging technical issues also present as E-commerce systems are complex, distributed applications with many diverse users, heterogeneous hardware, networking and user interface technologies, have important security, integrity and performance constraints, and are becoming "mission critical" parts of many organisations' IS Architectures (Diffie, 1998; Evans & Rogers,

1997; Jutla *et al*. 1999a; Pour 1999). Developers need to appropriately specify and design such complex applications, including supporting not only functional specifications but also the myriad of complex non-functional constraints. They also need to develop complex software architectures, good user interfaces (often for non-expert users), and implement systems using appropriate distributed system middleware, data management and user interface technologies. Integration with existing systems is almost always required, and can be challenging.

The quality of E-commerce applications is important to assure, as with any IS project, but presents many new challenges. Testing of highly distributed E-commerce applications is difficult, especially given the range of hardware and networks parts of the system may be deployed on. Usability of the product is crucial, particularly if customers or other business users need to interact with it (Singh *et al*. 1999). Ensuring an E-commerce system meets its usually complex non-functional constraints often takes considerable effort and expense. Finally, determining that an E-commerce system meets an organisation's staff, customer and related business needs, is in many ways still an open research topic.  While some traditional organisational effectiveness metrics can be used, market research analysis and other techniques increasingly need to be used.

When considering a course, set of courses or even a whole programme on E-commerce systems development, one should carefully determine appropriate Information systems educational techniques are appropriately chosen to ensure the range of issues described above are adequately addressed. Approaches can include using only lectures and small assignments, using project-based learning, to using only directed readings. Different teaching styles suit different materials and learning outcomes – conceptual and theoretical material often suits lecture-based courses; practical techniques, experience and group learning suits project- or problem-based learning; and in-depth intellectual critiquing often suits a directed reading-based course structure.

We have taught many IS development courses, using project-based learning as a central technique (Grundy 1996; Grundy, 1997), i.e. students learning by doing individual and group projects. This would also seem appropriate for much of E-commerce systems engineering education, given the practical skill set intended for graduates focusing on this area (Dhamija *et al*. 1999; Ge & Sun, 2000). In addition, as Information Technology used for E-commerce systems development changes so rapidly, it is crucial to ensure students are "educated for change" in this area, and become intellectually independent as a result of any course(s) they take (Grundy, 1996; Ross & Ruhleder, 1993). Students should be educated so they are able to compare and contrast new development methods, techniques and tools as they become aware of them while practising in industry, and be able to critically evaluate the many options at various levels they will have for developing E-commerce systems, to be able to make the best choices on their development projects. Given the rapid change in the Information Technology field, and in E-commerce systems engineering in particular, encouraging a desire among students to continue to learn and extend their skill set while in industry is essential.

## 3. COURSE OVERVIEW

Our graduate E-commerce Systems Engineering course is taught as part of the final year of an Honours degree programme, or as part of the first year of a Masters programme. The University of Waikato, where the course is taught, offers several specialised programmes as part of its Bachelor of Computing and Mathematical Sciences (BCMS) degree. The most popular is the Programme in Information Systems. The pre-requisite courses for our E-commerce course taken by students in this degree programme are illustrated in Figure 1. Our IS programme is organised to build up students skill set, conceptual understanding of IS Engineering issues and provide successively more advanced technical skills in a year-by-year fashion. Part 4 is an Honours year, though Part 4 courses can also be taken in the first year of a master of Computing and Mathematical Sciences (MCMS) degree.

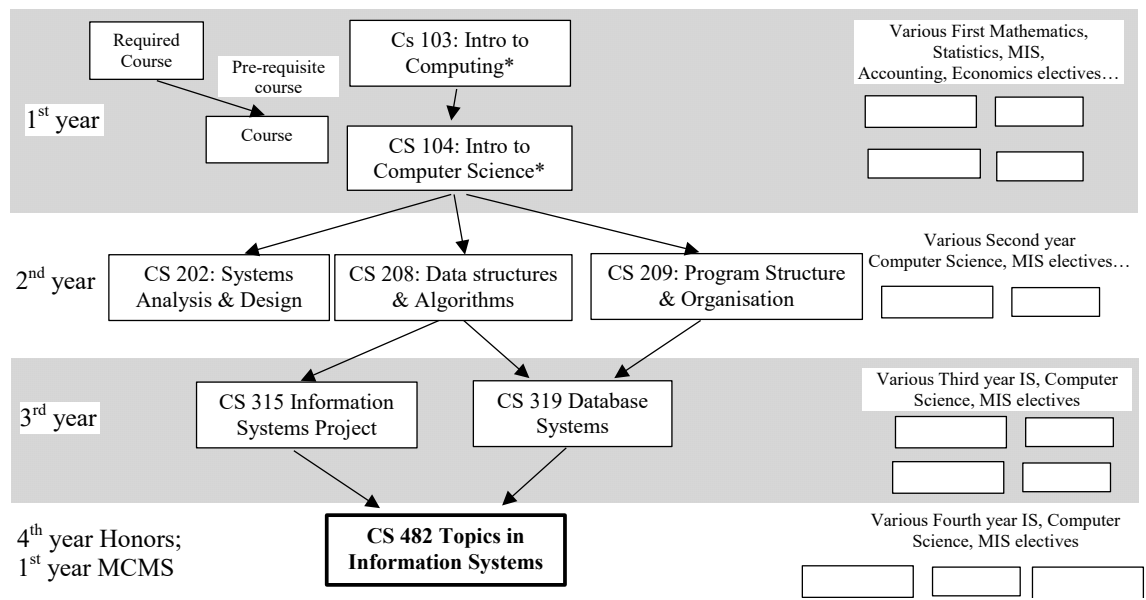

Figure 1. Structure of the BCMS in Information Systems Programme.

The "Topics in Information Systems" course, the subject of this paper and taught by the author in 1999 (and an equivalent course at the University of Auckland in 2000), was changed to emphasise E-commerce Systems Engineering. Most students taking Topics in Information Systems have done required Part 3 IS courses (e.g. IS Project, Databases, Network and Internet management), along with some MIS courses (e.g. Business Process Reengineering, Advanced Systems Analysis). The only specified pre-requisite is the IS Project course at Stage 3. This gives students experience in doing a substantial group IS development project in industry.

Most students taking Topics in Information Systems are BCMS IS programme majors, but the course also includes other BCMS majors and MCMS students, along with some MIS majors. Given these diverse backgrounds, we made the decision to give the course a strong practical IS Engineering focus. E-commerce systems were chosen as the target kinds of IS systems to study, given the huge demand for skills in this area from industry and strong desire to learn about the engineering of these systems of students. Also, they are usually one of the more challenging kinds of systems to determine requirements of, design, implement, evaluate and deploy. The key aim was to give all students a useful, balanced mix of E-commerce engineering theory and conceptual material, practical techniques for engineering E-commerce systems, and experience at working on a realistic, moderately-sized E-commerce engineering project.

To this end, the course curriculum was intended to cover issues pertinent to managing E-commerce Engineering projects, various application domains and deployment issues, Requirements Engineering, Software Architectures and Design, User interface design and implementation, distributed systems implementation using middleware technologies, IS evaluation approaches, and CASE tool support issues. This wide range of topics would build upon basic IS development skills students learned in the pre-requisite course, IS Project, as well as complement those from the Database, Internet management and various MIS courses. The delivery methods of the course would need to allow students from diverse backgrounds to not only gain a basic understanding of all of these key E-commerce engineering topics, but allow in-depth study of selected E-commerce Engineering topics that suit their own areas of specialisation.

The delivery methods chosen were also intended to be more holistic, breaking the traditional mould of lectures and assignments as used in most Auckland and Waikato University IS and CS courses. The class meets 3 times each week for 12 weeks, and lectures are limited to one per week over the 12 week semester, with interactive tutorial activities and simulations, practical lab workshops, student presentations, group meetings and individual research projects used.All course materials are published on the internet, all student projects (IS engineering and research projects) have their own web pages, and many Web-based example E-commerce systems and Web-based papers are referenced. A printed course manual is provided to students containing project specifications, required and recommended readings, tutorial and practical exercise descriptions, project management materials and other supplementary information.  Originally a web-only version of this was planned, but students preferred a conventional printed version which to take to lectures, tutorials and other meetings.

## 4. THEORY AND CONCEPTUAL MATERIAL

Topics in Information Systems takes a holistic view of E-commerce systems engineering, building on traditional IS Engineering material from prior IS courses. Its lecture content aims at providing a context for more interactive tutorials, practical exercises and project work, and for providing necessary theoretical content delivery and explanation in conjunction with selected readings. This combination of lectures as context and explanation with readings as background theoretical content allows us to appropriately emphasise practical techniques and experience, in comparison to many traditional Graduate IS courses that focus strongly on theory and conceptual material. Lectures and readings are used to guide students to appropriate conceptual material, and students are encouraged and expected to make critical judgements about such material. This is evidenced by critiques of the material in workbooks and student presentations. We also wanted to help students more carefully relate the practical material and group project work to the conceptual material in various ways. Evidence of making these conceptual links is expected in their workbooks, handed in at the end of the course. Thus we aim for a balanced mix of directed readings, lectures and problem-based learning, rather than emphasising one over the others.

Each week of Topics in Information Systems has a "theme" with a one hour lecture each week centred on this theme given by the course professor. One required reading related to the week's theme is set, and an additional "recommended" reading provided. Additionally, various library and web-based resources relating to the week's theme are given, although students are also expected to carry out their own research and locate relevant web-based or library-based material. Table 1 illustrates each week's theme in 1999 and 2000, along with required readings. Full details can be found off the course web-site: www.cs.auckland.ac.nz/415.702SC/.

| # | Theme | Reading | Presentation | Activity | Practical |
|---|---|---|---|---|---|
| 1 | Software Processes | The Personal Software Process, Humphrey 96, IEEE Software | Where are the silver bullets? | | Java work scheduler |
| 2 | E-commerce & project man. | Taxonomy of E-commerce, Bambury First Monday, 1998 | Virtual organisations | E-commerce simulation | Client-server scheduler |
| 3 | Requirements Engineering | Use Cases, Booch *et al*., UML User Guide | Formal requirements | Use cases & KAOS | MS Access forms |
| 4 | OOA (UML) | Class diagrams, Fowler, UML Distilled | OOA vs. SA | OOA with UML for E-commerce | Scheduler using Database |
| 5 | Software Architectures | Designing system architecture, Quatrani, UML & Rose, 1998 | Client-server Architectures | SA simulations | 3-tier scheduler |
| 6 | OOD (UML) | OO Design, Conger, The New Software Engineering, 1994 | Y2K design problem | SA & OOD Evaluations | Group Project Work |
| 7 | User Interfaces | Choosing UI development tool, Valer, IEEE Software, 1997 | Web-based IS Interfaces | E-commerce UIs | Group Project Work |
| 8 | Distributed systems technologies | Java Applets and CORBA, Evans, Internet Computing, 1997 | CORBA vs DCOM | ORB, RPC simulations | Group Project Work |
| 9 | CASE Tools | Classification of CASE, Fuggetta, COMPUTER 1993 | Rose evaluation | Round-trip engineering | Group Project Work |
| 10 | MetaCASE Tools | MetaEDIT+, Kelley, CAiSE'96 | Presentations | E-commerce tool design | Group Project Work |
| 11 | IS Evaluation | Basic evaluation techniques, Preece, HCI, 1994 | Presentations | IS Evaluation | Group Project Work |
| 12 | Future of ISE | Innovation & Obstacles, Clark, COMPUTER 1998 | Presentations | | Group Project Work |

Table 1. An outline of the course content (as used in 1999 and 2000) by week.

A wide range of E-commerce and IS engineering topics are covered and no one topic can be fully addressed in any great depth. This approach was deliberate, and allows students to gain basic understanding of important issues related to E-commerce systems engineering for each theme topic, but also allows them freedom to more fully explore selected issues on their own or in their project groups as is relevant to each individual and/or group. Different professors can "tune" material to suit their particular interests or adapt it to new directions emerging in the IS/E-commerce engineering areas. For example, the course could focus more on E-commerce IT policy, project management or requirements engineering, or on architecture and implementation, or on evaluation and tools as the professor feels is appropriate. As this course has a wide student intake from varied backgrounds, including IS Engineering, Computer Science and MIS majors, the broad topics help ensure all finish with understanding and experience of the main issues of E-commerce systems development, but are able to focus on areas of their own interest or relevance. The individual research topic provides this outlet, allowing students to focus in on a selected sub-topic and research it in depth. We prefer students to refer a research topic of relevance to them rather than set them for this reason. We found MIS majors focus on management and requirements issues, Computer Science majors on design methods and technologies.

## 5. PRACTICAL TECHNIQUES

Students participate in one one-hour tutorial per week, do five practical exercises during the course, and give one group "lecture" to the class as a group of 3-4. The aim of the tutorial is to "bring to life" the theme issue of each week in a joint problem-solving or group simulation exercise. Many of these focus on getting practice in groups with various IS engineering methods, while others focus on understanding problems, processes and technologies through interactive simulations. Table 2 shows the tutorials we have used to date. These, like lectures and readings, can be tuned to suit a particular professor's themes/expertise as necessary.

Student feedback has indicated these tutorials were one of the most liked and valuable parts of the course. The group problem-solving exercises greatly helped many students understand and apply key E-commerce systems concepts, and the small group approach enabled students to learn mainly through discussion and co-operation. The simulations were great fun for students and professor alike, as well as very illuminating. The simulations allow students to experience first-hand behavioural characteristics of E-commerce systems from various levels, as well as some non-functional characteristics. For example, when simulating E-commerce systems from a users' perspective, key interactions are discussed. When simulating from an architectural perspective, different components and their communication issues are investigated. Non-functional issues like quality of service (stopping students who are "networks" passing messages), performance (creating "bottle-necks" using tables for students acting as networks), and integration (having students translate "messages" in one format into another) become concrete in students' minds. Some of the more popular tutorials included the groupware systems simulation, which had students create a document simultaneously on a piece of paper (a "shared workspace"), and necessitated developing conflict resolution and work allocation/co-ordination approaches. The middleware tutorial involved passing "messages" over (sometimes faulty or slow) "networks" in paper cups, and included locating remote services, marshalling and de-marshalling information, as well as handling remote service and network failure. The CASE tutorial involved simulating

CASE tool reverse engineering and code generation to and from a CASE tool "repository", and a meta-CASE tutorial involved designing new tools for E-commerce systems development.

| # | Activity | Description |
|---|----------|-------------|
| 1 | E-commerce project simulation | Simulate an e-commerce system from both user and architectural viewpoints. Discuss issues to do with organising and running a project to develop a system. |
| 2 | Use cases & KAOS | Use UML Use cases and basic KAOS formal specifications to capture requirements about an example E-commerce system. Compare and contrast these approaches to codifying requirements. |
| 3 | OOA with UML | Develop in groups several UML diagrams to specify parts of an example E-commerce system. Compare and contrast different specifications developed. |
| 4 | SA simulations | Simulate several different E-commerce system architectures in groups. The architectures are quite different e.g. simple client-server vs decentralised vs web-based. |
| 5 | SA & OOD Evaluations | Given three example architectures for an E-commerce system specification, develop part of OOD model for system in groups. The architectures are from the 4$^{th}$ activity and have quite different designs. |
| 6 | Designing E-commerce user interfaces | Design a set of user interfaces for part of an example E-commerce system. Simulate example use of these interfaces, and in particular groupware interfaces. |
| 7 | ORB, RPC simulations | Simulate socket, database, RMI and CORBA-style distributed system implementations in groups. This involves passing "messages" between "systems" using the facilities such architectures provide. |
| 8 | Round-trip engineering | Simulate code generation and reverse engineering support as found in CASE tools. This involves generating "code" from a design, and after change to the "code", reverse engineering changes to the design. |
| 9 | Tool design | Design a CASE tool for a particular kind of E-commerce systems engineering domain. This involves identifying suitable tool features and discussing design and implementation issues for each. |
| 10 | IS Evaluation | Carry out a group evaluation of a given, real E-commerce system. A wide range of systems are investigated by different groups, including ATMs, telephone banking, on-line purchasing, on-line banking and virtual organisations. |

Table 2. An outline of the tutorial activities.

The practical exercises are used to familiarise students with implementation language, database and middleware technologies used in the group development project to build a prototype E-commerce system. In 1999 and 2000 this involved using MS Access forms and reports, an SQL server database, Java applets and servlets, and Java JDBC, sockets and RMI. This could easily be changed to using different technologies, such as Active Server Pages and CGIs, different database systems and user interface builders, and different middleware technologies and development environments. Table 3 shows the practical exercises used in 1999 for students who were not very familiar with Java but had used Access and SQL Server, and in 2000 for students experienced with Java but not so familiar with MS Access and SQL Server.

| #  | 1999 Practicals | 2000 Practicals |
|----|-----------------|-----------------|
| 1  | Introductory tutorial to Java language facilities. | Build a simple Java project management work scheduler application. |
| 2  | Build simple project management work scheduler tool using Java. Introduces students to main Java facilities. | Extend to a two-tier distributed scheduler (client and server) using Java RMI or sockets. |
| 3  | Extend to a simple client-server version of scheduler. Gives practice with sockets in Java. | Build MS Access forms to maintain scheduler database. This gives students practice using a database and basic Access forms. |
| 4  | Build an Access database with schedule items & Java access to database. Gives experience with JDBC. | Extend scheduler to access and update Database information directly. |
| 5  | Extend to a three-tier scheduler with applet, application server and database. These techniques used in project for building E-commerce systems. | Extend to use applet, application server and database i.e. a three-tier, scheduler. Techniques used in project to build E-commerce systems. |

Table 3. Practical exercises.

Student presentations (or "lectures") give students practice at researching, preparing and delivering oral presentations of technical material, as well as further group work experience. They also motivate students to learn in more depth issues relating to the theme issue of their presentation. Each presentation relates loosely to each week's theme issue, with students having considerable flexibility in the focus of their presentation and material they use. Most presentations were very professionally prepared and delivered, with most student groups using software demonstrations, PowerPoint presentations and other multi-media facilities like video. Table 4 shows a summary of student presentation topics for 1999.

| # | Presentation Topic | Material Presented by Students |
|---|--------------------|-------------------------------|
| 2 | Virtual Organisations | Group overviewed the issue of virtual organisations and their requirements for E-commerce systems support. Also briefly discussed issue of virtual software organisations. |
| 3 | Formal vs informal requirements specification | Group compared and contrasted UML Use case requirements modelling and KAOS formal RE modelling. |
| 4 | OOA vs Structured Analysis | Group compared and contrasted UML class diagrams and collaboration diagrams with SSAD dataflow and ER diagrams. |
| 5 | Client-server Software Architectures | Group overviewed current trends in client-server computing, and briefly discussed issues relating to technologies and architecture design. |
| 6 | Y2K Problem | Group overviewed y2k problem as a "design" problem. |
| 7 | Web-based E-commerce system user interfaces | Group discussed various design and implementation issues for web-based Information Systems, particularly E-commerce system user interfaces. |
| 8 | CORBA vs DCOM | Group compared and contrasted CORBA and DCOM distributed object technologies for building E-commerce systems. |
| 9 | Current CASE Tool Technology | Group surveyed commonly used CASE tool facilities and their suitability for supporting E-commerce systems engineering. |

Table 4. Student presentations.

## 6. PROJECT WORK

Two projects are carried out by students: one a group E-commerce systems engineering project, and the other an individual research project. The group development project gives students both group work experience and allows them to use the practical development techniques covered in lectures, tutorials and practicals on a moderately-sized realistic problem. The specification for the group project was deliberately open-ended: develop an E-commerce system for some organisation. Students had to ensure the system corresponded loosely to the architecture illustrated in Figure 2. The student's system needs to provide: various staff interfaces (maintain product/service details, generate reports, service customer on-line requests etc.); an on-line interface for customers (to search for and purchase products, check on order status, update information if required, etc.); and an interface to a "bank" credit card processing server. The bank server is provided and students have to incorporate electronic payment facilities with it.

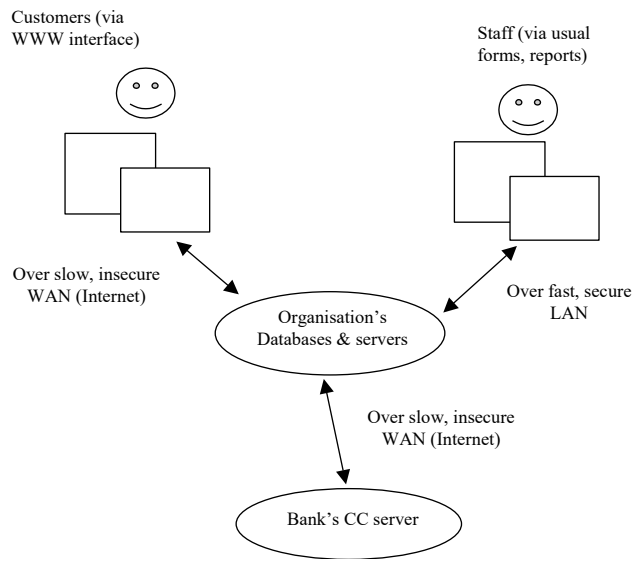

Figure 2. E-commerce systems architecture for the group development project.

The scope of the final project report was to describe the requirements of the particular E-commerce system the group has developed, a comprehensive specification, software architecture and design using the Unified Modelling Language, a prototype developed using MS Access™, SQL Server™, Java and HTML, and the results of an evaluation of the prototype. A wide range of projects were developed by students, as illustrated in Table 5.

| # | Some Group Projects | Some Individual Projects |
|---|---------------------|--------------------------|
| 1 | On-line book store. Similar concept to Amazon.com™. The student provided a rich set of navigation mechanisms and purchasing functions. | Visual Basic for E-commerce systems engineering. Compared using Visual Basic and active server pages to using HTML, CIGs, DHTML and Java applets/servlets for building E-commerce applications. |
| 2 | On-line clothing store. Similar to the book store, but the student used HTML and Java applet images to provide a very rich set of multi-media information, in addition to search-and-buy functionality. | CORBA for E-commerce systems infrastructure. Compared using a CORBA-based middleware infrastructure to using sockets, database connections (via SQL) and Java RMI for E-commerce systems. |
| 3 | Virtual software store. Provided a software sales site, using a brokering metaphor i.e. site allows customers to buy from various vendors via one web site. | Rapid Applications Development. Compared the RAD development process and tools to more traditional Information Systems development methods, especially for E-commerce systems development. |
| 4 | Furniture store. Used a concept of on-line sales to augment traditional sales. Customers could buy on-line but also browse store in person, or browse store items on-line and buy in-person. Nice mix of traditional and E-commerce retail concepts. | Enterprise Resource Planning Systems and E-commerce. Overviewed ERP system facilities and architectures, and discussed implementing E-commerce systems via such packaged software solutions. |
| 5 | Cookery site. Used a concept of "shareware" recipes and cooking materials, which was very interesting. Also has a very rich range of HTML/Images to augment searching, publishing and discussion functionality. | XML for E-commerce systems development. Gave an overview of the eXtensible Markup Language and discussed its possible role in E-commerce systems development, especially for Business-to-Business E-commerce applications. |
| 6 | Virtual art gallery. A rather novel system where artists publicise their works in a co-operative manner and customers purchase works. Nice combination of HTML and Applet interfaces. | Electronic money systems. Overviewed the concept of electronic money, compared and contrasted different approaches to supporting electronic money in E-commerce applications, and important security and integrity issues. |

Table 5. Some examples of group E-commerce system development projects.

The groups keep a log of meetings, tasks assigned to and deliverable work done by each member, and timesheets. Some process improvement information, such as defects detected and corrected at each stage of development, were also kept. We allowed groups to self-select into 4-5 people. Milestones include draft Requirements, specification and design and prototype, in addition to a final project report. The scope and quality of the developed E-commerce system prototypes has been quite astounding. Even groups containing mainly MIS majors who has significantly less software development experience than IS and CS majors produced excellent, working systems. Most of the systems outlined in

| # | Some Group Projects | Some Individual Projects |
|---|---------------------|--------------------------|

| 1 | On-line book store. Similar concept to Amazon.com™. The student provided a rich set of navigation mechanisms and purchasing functions. | Visual Basic for E-commerce systems engineering. Compared using Visual Basic and active server pages to using HTML, CIGs, DHTML and Java applets/servlets for building E-commerce applications. |
|---|---|---|
| 2 | On-line clothing store. Similar to the book store, but the student used HTML and Java applet images to provide a very rich set of multi-media information, in addition to search-and-buy functionality. | CORBA for E-commerce systems infrastructure. Compared using a CORBA-based middleware infrastructure to using sockets, database connections (via SQL) and Java RMI for E-commerce systems. |
| 3 | Virtual software store. Provided a software sales site, using a brokering metaphor i.e. site allows customers to buy from various vendors via one web site. | Rapid Applications Development. Compared the RAD development process and tools to more traditional Information Systems development methods, especially for E-commerce systems development. |
| 4 | Furniture store. Used a concept of on-line sales to augment traditional sales. Customers could buy on-line but also browse store in person, or browse store items on-line and buy in-person. Nice mix of traditional and E-commerce retail concepts. | Enterprise Resource Planning Systems and E-commerce. Overviewed ERP system facilities and architectures, and discussed implementing E-commerce systems via such packaged software solutions. |
| 5 | Cookery site. Used a concept of "shareware" recipes and cooking materials, which was very interesting. Also has a very rich range of HTML/Images to augment searching, publishing and discussion functionality. | XML for E-commerce systems development. Gave an overview of the eXtensible Markup Language and discussed its possible role in E-commerce systems development, especially for Business-to-Business E-commerce applications. |
| 6 | Virtual art gallery. A rather novel system where artists publicise their works in a co-operative manner and customers purchase works. Nice combination of HTML and Applet interfaces. | Electronic money systems. Overviewed the concept of electronic money, compared and contrasted different approaches to supporting electronic money in E-commerce applications, and important security and integrity issues. |

Table 5 were of close to commercial quality in terms of scope, functionality, usability and the potential for extension.

The individual research project gives students an in-depth understanding of a particular area of E-commerce systems development. There are no intermediate deliverables except an initial one page proposal early in the course. The project specification is to write a 6 to 8 page paper (in IEEE Press format), on a selected IS engineering topic. In addition, students need to write a set of web pages describing their project and provide additional information about their paper topic, including useful web resources, and give a short (15 minute) conference-style presentation of their paper to the class. Table 5 shows some of the topics covered in individual research projects.

## 7. ASSESSMENT

Assessment is made up of four equal parts: group development project; individual research project; final examination; and student workbook. Students are required to keep a workbook throughout the course. Their workbook mark is based on a combination of: evaluation of group class presentation; tutorial and practical exercise write-ups; and critiques of various course materials (lectures, readings, web sites, other group presentations etc). Encouraging students to critique the course lectures and web-site, as well as other group presentations and readings, has proved popular, giving students a chance to formally give structured feedback on the course.

The group project is assessed as an equal mark for all group members, based on the final report and prototype software. A small variation in this mark per group member is allowed for based on anonymous peer assessments handed to the lecturer at the end of the course reporting each group member's assessment of their own and other's performances on the project. This makes group assessment straightforward and reasonably fair (Grundy, 1997). The individual project paper is "refereed" by the course professor, according to usual IS conference refereeing criteria, and mark for this project comprises the paper "result" plus web site and class paper presentation assessment. We found students quite enjoy the concept of writing a "paper" and "submitting" it for "refereeing", and having in-class presentations treated like a conference paper presentation.

The examination is a two hour essay on one selected E-commerce systems engineering topic, chosen by students from a small selection. Essays should demonstrate a good understanding of the chosen topics, can use appropriate references to course readings or other materials to back up their statements, and should produce a coherent argument for their point of view. The essay examination is open book, and thus students have available all their notes as well as the course manual with copies of the required and recommended readings. The idea of the essay examination is thus not to test recall ability, but to test the ability of students to formulate a coherent argument relating to their selected question, drawing on appropriate course resources.

## 8. DISCUSSION

Topics in Information Systems has succeeded in delivering a balanced mix of theoretical and conceptual material, practical IS development methods and techniques, and project experience in developing E-commerce systems, for a diverse range of students. Giving each week of the course a theme ensures students gain a basic understanding of a range of important E-commerce engineering issues, as well as helping to structure course pace and focus. The individual research project, readings and lectures, and group development project all allow scope for in-depth study of selected E-commerce engineering topics. We were pleased that a balance of MIS, IS Engineering and Computer Science material was covered during the course, all tailored to focus on E-commerce systems engineering. This made the course attractive to a wide range of students, and this range of student backgrounds proved to beneficial to all involved, providing a rich, co-operative learning environment throughout the course. E-commerce systems have proved a very popular focus for students, and are an excellent example of challenging IS development.

Particularly successful parts of Topics in Information Systems proved to be the tutorials, group project work and student presentations. Tutorials which emphasise group problem-solving with new techniques, or interactive group simulations, were a lot of fun as well as being very educational. Working in groups was found very useful by students, despite many already having considerable group work experience from previous courses. Student presentations were of excellent quality and very professionally developed and delivered. Many employers have expressed concern to us in the past about graduates' poor written and oral presentation skills, and we were pleased to be able to give students guidance and practice in these important areas. The workbooks of most students proved to be of a high quality, although many found it difficult to compare and contrast methods and technologies, and to critique readings and presentations. Sadly, many students reported it was the first time they had been asked to do "critical comparison and analysis" tasks in their degrees!

The individual research project received good feedback from students in terms of the freedom to chose their own topic and emphasis, and its use of the concept of papers and paper presentations. Unfortunately, apart from a few outstanding examples, most of these projects were poor quality. Many papers were extremely poorly structured (despite the required and recommended readings giving students many examples of "good" papers!), many involved quite cursory "research" work, the use of references was almost non-existent, and the web pages developed by students usually just an electronic form of their paper. This was disappointing, and in future we plan to give students more guidance and coaching in writing technical papers.

The essay examination was not well-received by students, many claiming they had never had to write an essay before during their time at University (which seemed, to the course professor, an excellent reason why they should have at least one experience of this!). The essay topics were quite broad, and all asked students to compare and contrast various conceptual and practical E-commerce development issues. Students were advised to treat the essay as a "report" an industry IS manager may ask them to write to help them understand and compare various IS engineering issues. Despite a cool student reception to the essay concept, the essays on the whole were very good, several being of exceptional quality.

Formal and informal student evaluations of Topics in Information Systems have been excellent. The general consensus has been that the course in its form described in this paper provides an excellent combination of breadth and depth, with students able to individually gain additional expertise in self-selected areas of E-commerce systems engineering. Particularly popular aspects include the tutorials, group and individual projects and the weekly themes. The focus on E-commerce systems engineering was very popular. Several students thought the prototype systems they developed were of near-commercial quality, an opinion shared by the course professor. Many students used their project reports as additional material when applying for industry jobs. Feedback from employers requesting references for students from the course professor have been very positive about the course and student expertise in E-commerce systems engineering.

## 9. SUMMARY

Our graduate-level course Topics in Information Systems has been modified to focus on E-commerce Systems Engineering theory and concepts, practical techniques and practical experience. A range of E-commerce engineering topics are covered, with delivery via lectures, student presentations, readings, web resources, tutorials and practical exercises. In addition, two projects, one a prototype E-commerce system development and the other an individual research project, give students experience in using concepts and techniques presented in the course. This has been a very successful course in terms of the quality of development projects, the interest level of students and in both student and industry feedback. The model for an advanced E-commerce Systems engineering course as presented in this paper has worked very well in the context it has been used: augmenting fairly traditional four-year Information Systems Engineering honours programme and Masters degree. Particular features are its holistic approach to teaching E-commerce systems engineering topics and its emphasis on student-directed learning. We believe the model for this course would also work well if more E-commerce related courses were taken by students prior to this course, with appropriate modification of focus.